\documentclass[10pt,twocolumn]{paper}
\usepackage{epsfig, graphicx}
\usepackage[english]{babel}
\usepackage[margin=1.5cm]{geometry}

\title{\center \rm \bf On the Diffuse Structure of the Toluene -- Water Interface}

\author{\small \rm Aleksey M. Tikhonov$^{a,b}$\/\thanks{tikhonov@kapitza.ras.ru}\\ \small
Kapitza Institute for Physical Problems, Russian Academy of Sciences, Moscow, 119334 Russia\\
\small
Institute of Solid State Physics, Russian Academy of Sciences, Chernogolovka, 
Moscow region, 142432 Russia
}

\begin{document}
\maketitle
%\centerline{\today}

\abstract{ \it \normalsize The electric density profile along the normal to the phase interface between aromatic hydrocarbon toluene and water has been studied by X-ray reflectometry using synchrotron radiation. According to the experimental data, the width of the interface under normal conditions is ($5.7 \pm 0.2$)\,\AA{}. This value is much larger than a theoretical value of ($3.9\pm 0.1$)\,\AA{} predicted by the theory of capillary waves with an interphase tension of ($36.0 \pm 0.1$)\,mN/m. The observed broadening of the interface is attributed to its own diffuse near-surface structure with a width no less than $4$\,\AA{}, which is about the value previously discussed for high-molecular-weight n-alkane -- water and 1,2-dichloroethane -- water interfaces.
}

\vspace{0.25in}
%\large
\normalsize

Two forms of the transverse structure of the oil -- water interface that imply its fundamentally different
molecular structures, which affect, e.g., the rates of interphase ion exchange or chemical reactions, are
widely discussed [1-4]. This is either an interface with a clear molecular structure with the width determined only by roughness caused by capillary waves or an interface with a diffuse structure at which mixing of phases occurs at a molecular level in a layer with a thickness up to several molecular dimensions. 
In this work, the first measurement of the reflection coefficient from the interface between water and
aromatic hydrocarbon toluene (C$_7$H$_8$), which is considered as a model when studying, e.g., adsorption of asphaltenes [5], is reported. The found value of the interface width under normal conditions is significantly different from the prediction of the theory of capillary waves, which is unambiguously determined by the interphase tension $\gamma$ and temperature $T$. The observed smearing of the interface is attributed to its own near-surface structure, e.g., a phase mixing region with the width $\geq 4$\,\AA{}.

The planar toluene -- water interface oriented by the gravitational force was studied in a thermostated stainless steel cell ($T\approx 298$\,K) with X-ray transparent windows made from polyethylene glycolterephthalate using the method described in [6]. Aromatic hydrocarbon toluene 
(C$_7$H$_{8}$, density $\approx 0.86$\,g/cm$^3$ ïðè 298\,Ê, boiling temperature $T_b\approx 384$\,Ê)
was purchased from Sigma-Aldrich and was purified by multiple filtration in a chromatographic column [7]. Toluene and water hardly mix under normal conditions.

\begin{figure}
\vspace{0.5in}
\hspace{0.15in}
\epsfig{file=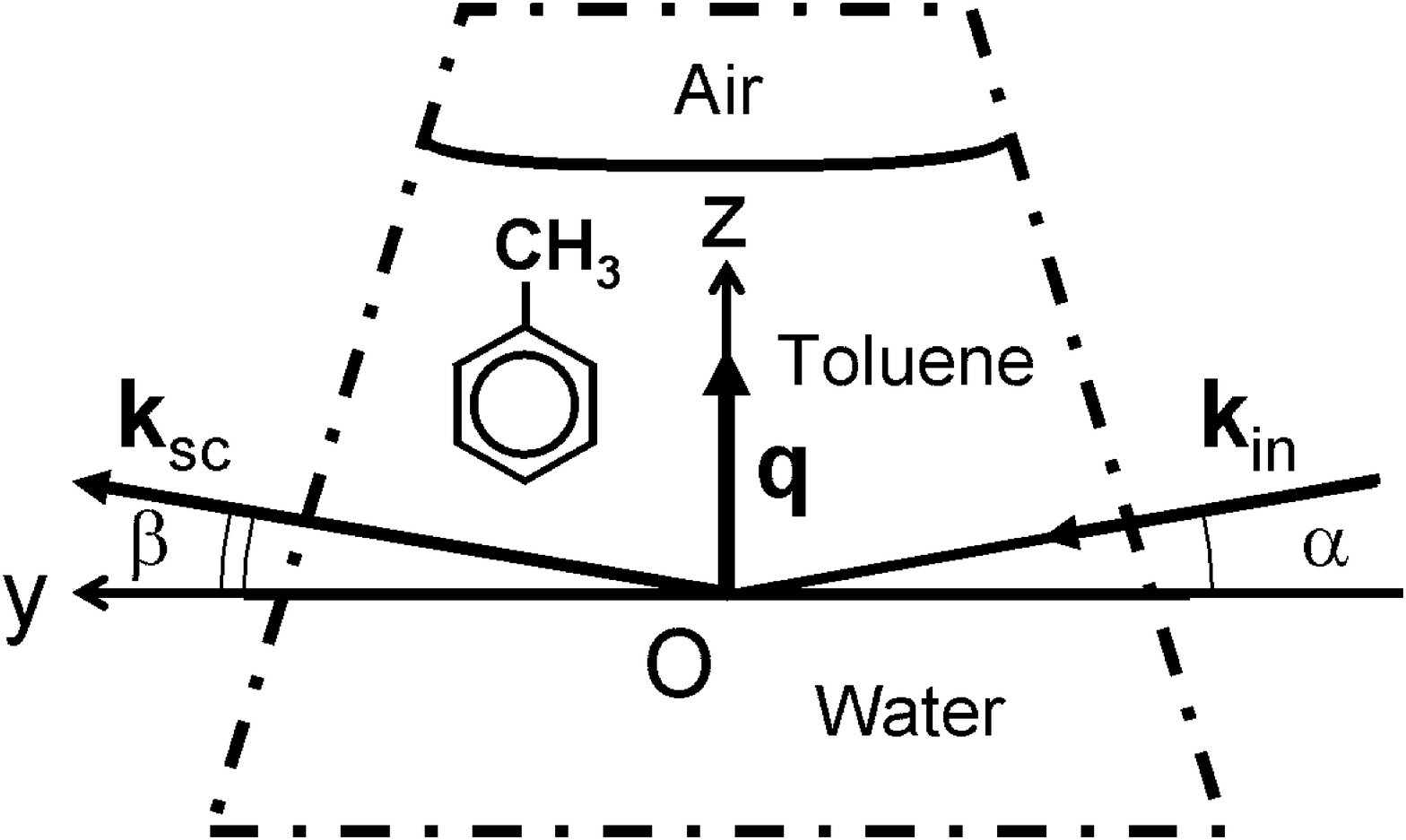, width=0.4\textwidth}

\vspace{0.15in}
\small {\bf Figure 1.} \it Kinematics of the X-ray surface scattering at the
toluene-water interface. In the reflectometry experiment, $\alpha = \beta$.
\end{figure}

The surface tension of the pure interface $\gamma$ was measured by the Wilhelmy plate method directly in
the experimental cell and was $\gamma = 36.0 \pm 0.1$\,mN/m, which is in good agreement with the data reported in [8-10]. About 100\,mL of deionized water with pH\,$\approx 7$ (Barnstead, NanoPureUV) was used as the lower bulk phase. About 50\,mL of toluene was used as the upper bulk phase. Before the measurements of the reflection coefficient $R$, the sample was "annealed" and was aged for several hours [11].

The transverse structure of the toluene -- water interface was studied by X-ray reflectometry in the X19C
beamline of the National Synchrotron Light Source. This beamline allows the study of both the surface of
solids and liquids and hidden liquid -- liquid interfaces [12-16]. A focused monochromatic beam with an
intensity of $\approx 10^{11}$\,photons/s and a photon energy of $E=15$\,keV 
(wavelength $\lambda=0.825 \pm 0.002$ \,\AA) was used in the experiments. This beamline was previously used to study phase transitions in adsorption layers of fatty alcohols and acids at the (saturated hydrocarbon–water) interface [11, 17].

Figure 1 illustrates the kinematics of the surface scattering at the interface, where $\alpha$ is the glancing angle and $\beta$ is the angle between the interface plane and the direction to the detector in the plane of incidence $yz$. In the reflectometry experiment ($\alpha=\beta$), X-rays pass through the oil phase and are reflected from the near-surface structure at the interface. In this case, the scattering vector {\bf q = k$_{\rm in}$ {\rm -} k$_{\rm sc}$}, where {\bf k}$_{\rm in}$ and {\bf k}$_{\rm sc}$ are the wave vectors of the incident and scattered beam in the direction of the observation point, respectively, is strictly perpendicular to the surface along the $Oz$ axis, i.e., in the direction opposite to the gravitational force. The measurement of the reflection coefficient $R$ as a function of $q_z=(4\pi/\lambda)\sin\alpha$ allows probing the transverse structure of the interface. The $R(q_z)$ values in the experiment are averaged over a macroscopic area of $\sim 0.5$\,cm$^2$ because of the geometrical dimensions of the incident beam ($> 5$\,$\mu$\,m and $\sim 2$\,mm in the vertical and horizontal planes, respectively).

\begin{figure}
\hspace{0.15in}
\epsfig{file=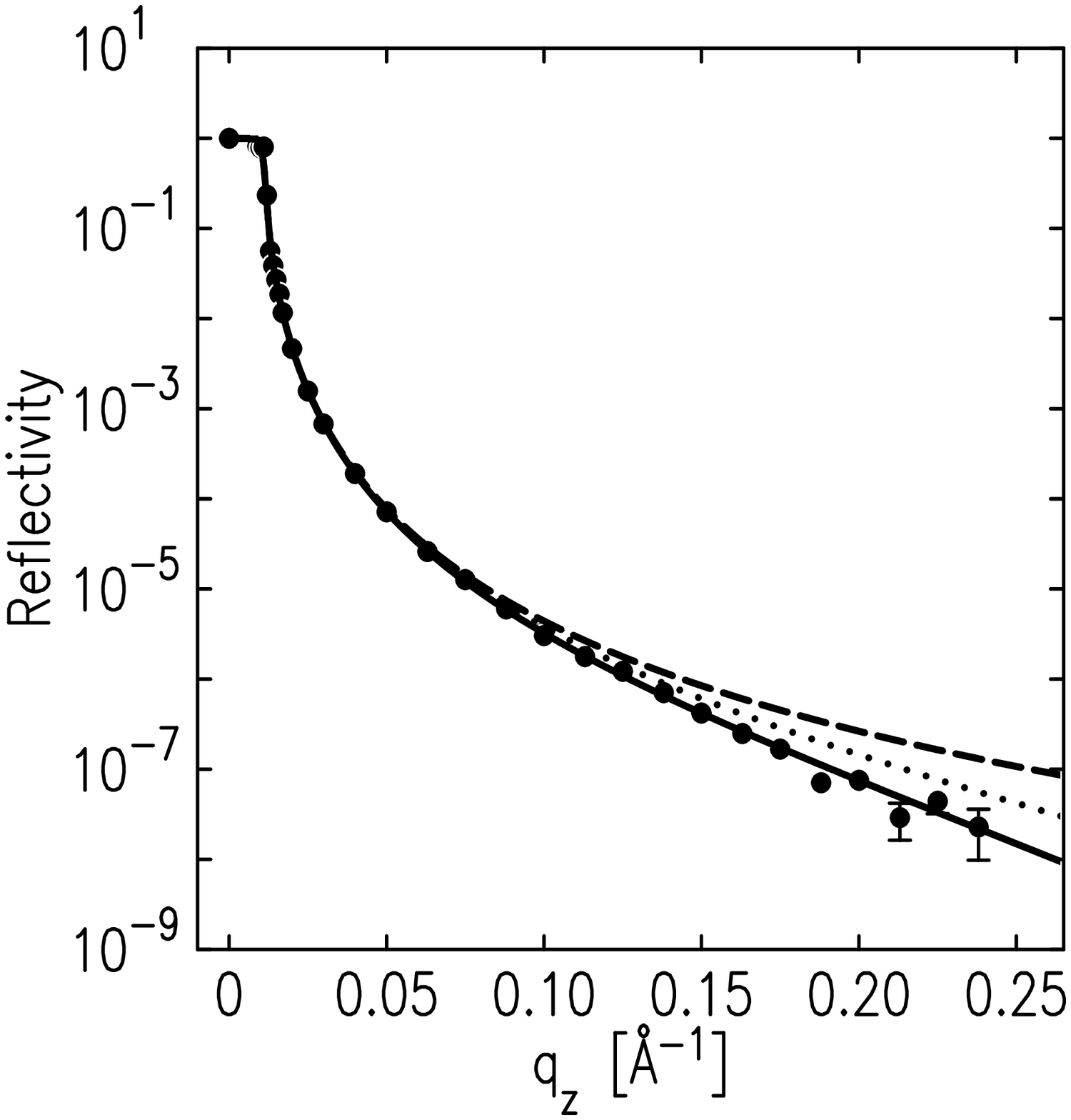, width=0.45\textwidth}

\small {\bf Figure 2.} \it X-ray reflection coefficient $R$ for the toluene -- water interface versus $q_z$. The dashed line is the Fresnel function (5), the dotted line is the calculation by Eq. (4)
with $\sigma=\sigma_{cw}$, and the solid line is the one-parameter model (1) with the fitting parameter ($\sigma = 5.7\pm0.2$)\,\AA.

\end{figure}

The electron number density under normal conditions is $\rho_w\approx 0.333$ {\it e$^-$/}{\AA}$^3$
in water and $\rho_{t} \approx 0.86 \rho_w$ in toluene. At glancing angles smaller than $\alpha_c\approx\lambda\sqrt{r_e(\rho_w - \rho_t)/\pi}$$\approx 6 \cdot10^{-4}$\,rad (where $r_e = 2.814\cdot10^{-5}$ {\AA} is the classical radius of the electron), the incident beam undergoes the total external reflection; i.e., $R\approx 1$. The angle $\alpha_c\approx 0.03^\circ$ is approximately a factor of 1.5 smaller than that for the previously studied n-hexane -- water interfaces [18, 19].

The experimental dependence of the reflection coefficient $R$ on $q_z$ for the toluene -- water interface is shown by dots in Fig. 2. From these data, the electron density distribution along the normal to the
surface is reconstructed within the standard one-parameter model using the error function [20-22]:
\begin{equation}
\begin{array}{l}
\displaystyle
\langle \rho(z) \rangle =\frac{1}{2}(\rho_w+\rho_t)+\frac{1}{2}(\rho_w-\rho_t){\rm erf}\left(\frac{z}{\sigma\sqrt{2}}\right),
\\ \\
\displaystyle
{\rm erf}(x)=\frac{2}{\sqrt{\pi}}\int_0^x\exp(-y^2)dy,
\end{array}
\end{equation}
where $\sigma$ is the rms deviation of the interface from the nominal position $z=0$.

Within the hybrid model, 
\begin{equation}
\sigma^2=\sigma^2_{0} + \sigma^2_{cw},
\end{equation}
where $\sigma_{0}$ is the proper width of a noncapillary-wave nature. The estimate of the integral characteristic of the spectrum of capillary waves for the isotropic surface of the liquid or the "capillary width" squared $\sigma^2_{cw}$ is determined by the range of surface spatial frequencies
covered in the experiment [23-25]:
\begin{equation}
\sigma_{cw}^2 =  \frac{k_BT}{2\pi\gamma} \ln\left(\frac{Q_{max}}{Q_{min}}\right).
\end{equation}
Here, $Q_{max} = 2\pi/a$ is the short-wave-length limit of the spectrum ( $a\approx 10$ {\AA} is the intermolecular distance in order of magnitude) and is the long-wave-length limit, where $\Delta\beta$$\approx 4\cdot10^{-4}$\,rad is the angular resolution of the detector in the experiment and $q_z^{max} \approx 0.25$ {\AA}$^{-1}$. Consequently, Eq. (3) for the toluene -- water interface in this experiment gives $\sigma_{cw} = 3.9\pm0.1$\,\AA.

In the first distorted wave Born approximation, the reflection coefficient has the form [26, 27]
\begin{equation}
 R(q_z)=R_F(q_z)\exp\left(-\sigma^2q_z\sqrt{q_z^2-q_c^2}\right),
\end{equation}
where
\begin{equation}
R_F(q_z)=\left(\frac{q_z-\sqrt{q_z^2-q_c^2}}{q_z+\sqrt{q_z^2-q_c^2}}\right)^2,
\end{equation}
is the Fresnel function and $q_c=(4\pi/\lambda)\sin\alpha_c$. 

In Fig. 2, the solid line is the one-parameter model (1) with the fitting parameter $\sigma = 5.7\pm0.2$\,\AA{} in (4), the dashed line is the Fresnel function (5), and the dotted line is the calculation by Eq. (4) with $\sigma=\sigma_{cw}$.

Thus, according to the experimental data, the electron density profile (1) for the toluene–water interface
has the width $\sigma=5.7 \pm 0.2$\,\AA{}, which is much larger than the value $\sigma_{cw}=3.9\pm 0.1$\,\AA{} calculated within the theory of capillary waves for the measured interphase tension ($36.0\pm0.1$)\,mN/m. This new and quite surprising experimental result indicates the existence of the structure of the interface with the width no less than $\sigma_0=\sqrt{\sigma^2-\sigma_{cw}^2}\approx 4.2$\,\AA{}, i.e., no less than the radius of the benzene ring of a toluene molecule ($\sim 4$\,\AA).

The previous reflectometry data indicated the absence of any diffuse layer in the near-surface structure
of the n-hexane -- water, nitrobenzene -- water, and 2-heptanone -- water interfaces [28-30]. However, the
studies of the n-alkane -- water and silica hydrosol -- air interfaces demonstrated that their interphase width can be described only with the inclusion of contributions both from capillary waves and from their own structure [3, 31]. In the former case, the width of their own structure is determined by two physically significant parameters of the system - the radius of inertia of a hydrocarbon molecule and the bulk correlation length [32, 33]. The latter parameter specifies, e.g., the width of the n-docosane -- water interface [6]. In the considered case, the observed width of the diffuse phase mixing region ($\geq 4$\,\AA) has the same order of magnitude as the width previously discussed for the n-hexadecane -- water and 1,2-dichloroethane -- water interfaces [3, 4].

To summarize, this study of the interface between aromatic hydrocarbon and water, as well as our previous
reports on phase transitions at saturated hydrocarbon-water interfaces, has demonstrated new experimental
capabilities provided by X-ray scattering methods with synchrotron radiation for revealing the
essence of processes occurring at phase interfaces in water -- oil emulsions in the presence of impurity surfactants (asphaltenes, naphthenic acids, etc.), which affect the efficiency of oil technological processes [34-36].

This work was performed using the resources of the National Synchrotron Light Source, US Department
of Energy (DOE) Office of Science User Facility, operated for the DOE Office of Science by the
Brookhaven National Laboratory under contract no. DE-AC02-98CH10886. The X19C beamline was supported
by the ChemMatCARS National Synchrotron Resource, University of Chicago, University of Illinois
at Chicago, and Stony Brook University. The theoretical part of the work was supported by the Russian Science Foundation (project no. 18-12-00108).

\end{document}